\documentclass[reprint,aps,pre,superscriptaddress]{revtex4-1}

\usepackage{textcomp}
\usepackage{amsmath}
\usepackage{amssymb}
\usepackage{mathrsfs}
\usepackage[retainorgcmds]{IEEEtrantools}

\usepackage{color, soul}
\usepackage{empheq}
 
\usepackage{graphicx}

\begin{document}


\title{Collective dynamics in atomistic models with coupled translational and spin degrees of freedom}


\author{Dilina Perera}
\email[]{dilinanp@physast.uga.edu}
\affiliation{Center for Simulational Physics, The University of Georgia, Athens, Georgia 30602, USA}
\affiliation{Department of Physics and Astronomy, Mississippi State University, Mississippi State, Mississippi 39762, USA}

\author{Don M. Nicholson}
\affiliation{University of North Carolina at Asheville, Asheville, North Carolina 28804, USA}

\author{Markus Eisenbach}
\affiliation{Oak Ridge National Laboratory, Oak Ridge, Tennessee 37831, USA}

\author{G. Malcolm Stocks}
\affiliation{Oak Ridge National Laboratory, Oak Ridge, Tennessee 37831, USA}

\author{David P. Landau}
\affiliation{Center for Simulational Physics, The University of Georgia, Athens, Georgia 30602, USA}



\begin{abstract}
Using an atomistic model that simultaneously treats the dynamics of translational and spin degrees of freedom,
we perform combined molecular and spin dynamics simulations 
to investigate the mutual influence of the phonons and magnons on their respective frequency spectra and lifetimes in ferromagnetic bcc iron.
By calculating the Fourier transforms of the space- and time-displaced correlation functions, 
the characteristic frequencies and the linewidths of the vibrational and magnetic excitation modes were determined.
Comparison of the results with that of the standalone molecular dynamics and spin dynamics simulations
reveal that the dynamic interplay between the phonons and magnons leads to a shift in the respective frequency spectra 
and a decrease in the lifetimes.
Moreover, in the presence of lattice vibrations, additional longitudinal magnetic excitations were observed with the same frequencies as the longitudinal phonons.
\end{abstract}


\maketitle


\section{Introduction}

For decades, dynamical simulations of atomistic models have played a pivotal role in the study of collective phenomena in materials at finite temperatures.
Molecular dynamics (MD)~\cite{Rapaport2004, Frenkel2002} utilizing empirical potentials 
has been extensively used in the analysis of vibrational properties in a variety of systems such as
metals and alloys~\cite{metal1, metal2, metal3, alloy1, alloy2}, polymers~\cite{polymer1, polymer2}, carbon nanotubes~\cite{nanotubes}, graphene~\cite{graphene} etc.
With regard to magnetic excitations, the lesser-known spin dynamics (SD) method~\cite{Landau1999, Tsai2000, Chen1994, Gerling1990, Blume}
has proven to be an indispensable tool for investigating classical lattice-based spin  models for which the analytical solutions are intractable.
Over the years, SD simulations have expanded our understanding of spin waves and solitons in magnetic materials,
leading to a number of groundbreaking discoveries, 
including the existence of propagating spin waves in paramagnetic bcc iron~\cite{Tao2005}, 
presence of longitudinal two-spin-wave modes~\cite{Bunker2000} that subsequently lead to experimental verification~\cite{TwoSpinWaves},
and an unexpected form of transverse spin wave excitations in antiferromagnetic nanofilms~\cite{Zhuofei2015}. 

Study of collective dynamics in magnetic materials faces an enormous challenge 
due to the coupling of lattice vibrations and spin waves which is inherently neglected in the aforementioned atomistic models.
In magnetic metals and alloys, the atomic magnetic moments and exchange interactions 
strongly depend on the local atomic environment~\cite{VariableMoments1, VariableMoments2, VariableMoments3} 
and therefore change dynamically as the local crystal structure is distorted by lattice vibrations~\cite{Yin2012}.
On the other hand, magnetic interactions themselves are integral for maintaining the structural stability of such systems~\cite{PhaseStabilityFeCo, Ekman1998}.
For instance, the stabilization of the bcc crystal structure in iron is long conceived to be of magnetic origin~\cite{PhaseStabilityIron, Herper1999}.
Furthermore, a number of recent studies emphasize the significance of phonon-magnon coupling on various dynamical processes such as 
self diffusion~\cite{SelfDiffusion}, thermal transport~\cite{ThermalTransport}, dislocation dynamics~\cite{DislocationDynamics}, and spin-Seebeck effect~\cite{Seebeck}. 
The dynamics of atomic and magnetic degrees of freedom are, hence, inseparable and should be treated in a self-consistent manner.

The idea of integrating spin dynamics with molecular dynamics was pioneered by Omelyan \textit{et al.}~\cite{Omelyan2001} in the context of a simple model for ferrofluids.
The foundation of this combined molecular and spin dynamics (MD-SD) approach lies in the unification of an atomistic potential and a Heisenberg spin Hamiltonian,
with the coupling between the atomic and spin subsystems established via a coordinate-dependent exchange interaction. 
With the use of an empirical many-body potential and a parameterized exchange interaction,
Ma \textit{et al.}~\cite{Ma2008} further extended MD-SD into a framework for realistic modeling of bcc iron.
The  parameterization developed by Ma \textit{et al.}~\cite{Ma2008} 
has since been successfully adopted to investigate various phenomena in bcc iron such as
magneto-volume effects~\cite{MagnetoVolumeEffects},
vacancy formation and migration~\cite{WenVacancy1, WenVacancy2},
and external magnetic field effects~\cite{MagField}.
Moreover, the method has been recently extended by incorporating spin-orbit interactions
to facilitate the dynamic exchange of angular momentum between the lattice and spin subsystems~\cite{spin_orbit}.
This, in particular, extends the applicability of MD-SD to accurate modeling of non-equilibrium processes. 

The aim of this paper is to improve our understanding of phonon-magnon interactions in the ferromagnetic phase of bcc iron within the context of MD-SD. 
This study is an extension of our earlier preliminary work~\cite{Perera2014, PereraST2014} 
which primarily focused on the effect of lattice vibrations on the spin-spin dynamic structure factor in the [100] lattice direction.
In this paper, we provide a more in-depth analysis of the mutual influence of phonons and magnons on their respective frequency spectra and lifetimes
for all three high-symmetry lattice directions: [100], [110] and [111]. 
This is achieved by comparing the results obtained for MD-SD simulations with those of standalone MD and SD simulations 
in which spin-lattice coupling is completely neglected.
In Sec.~\ref{sec:methods}, we present the MD-SD formalism and the parameterization for bcc iron, followed by a comprehensive description of the methods we adopt
for characterizing collective excitations.
Sec.~\ref{sec:results_1} and ~\ref{sec:results_2}, respectively, report our results on vibrational and magnetic excitations, 
followed by conclusions in Sec.~\ref{sec:summary}.

\section{Methods} \label{sec:methods}

\subsection{Combined molecular and spin dynamics}

MD-SD is essentially a reformulation of the MD approach,
in which the effective spin angular momenta of the atoms $\{\mathbf{S}_i\}$ are incorporated into the Hamiltonian
and treated as explicit phase variables.
For a classical system of $N$ magnetic atoms of mass $m$ described by their positions $\{\mathbf{r}_i\}$, 
velocities $\{\mathbf{v}_i\}$, and the atomic spins $\{\mathbf{S}_i\}$,
the MD-SD Hamiltonian takes the form
\begin{equation} \label{eq:hamiltonian}
  	\mathcal{H} = \sum_{i=1}^N \frac{ m{\mathrm{v}_i}^2 }{2} + U(\{\mathbf{r}_i\}) - \sum_{i<j} J_{ij}(\left\{ \mathbf{r}_k \right\}) \mathbf{S}_i \cdot \mathbf{S}_j,
\end{equation}
where the first term represents the kinetic energy of the atoms, 
and $U(\{\mathbf{r}_i\})$ is the spin-independent (non-magnetic) scalar interaction between the atoms.
The Heisenberg-like exchange interaction with the coordinate-dependent exchange parameter and $J_{ij}(\left\{ \mathbf{r}_k \right\})$ 
specifies the exchange coupling between the $i$th and $j$th spins.
The aforementioned Hamiltonian has true dynamics as described by the classical equations of motion
\begin{subequations} \label{eq:eom}
\begin{empheq}{align}
\frac{d\mathbf{r}_i}{dt} &= \mathbf{v}_i 							\label{eq:eom_pos} \\[4pt]
\frac{d\mathbf{v}_i}{dt} &= \frac{\mathbf{f}_i}{m}						\label{eq:eom_vel} \\[4pt]
\frac{d\mathbf{S}_i}{dt} &= \frac{1}{\hbar{}} \mathbf{H}_i^\text{eff} \times \mathbf{S}_i	\label{eq:eom_spin}
\end{empheq}
\end{subequations}
where $\mathbf{f}_i = -\nabla{}_{\mathbf{r}_i} \mathcal{H}$ and $\mathbf{H}_i^\text{eff} =\nabla{}_{\mathbf{S}_i} \mathcal{H}$ 
are the interatomic force and the effective field acting on the $i$th atom/spin.
The goal of the MD-SD approach is to numerically solve the above equations of motion
starting from a given initial configuration,
and obtain the trajectories of both the atomic and spin degrees of freedom.

MD-SD is a generic framework that with proper parameterization, can be readily adopted for any magnetic material 
in which the spin interactions can be modeled classically.  
In this study, we adopt the parameterization introduced by Ma \textit{et al.}~\cite{Ma2008} for bcc iron,
in which $U(\{\mathbf{r}_i\})$ is constructed as
\begin{equation} \label{eq:atomic_potential}
	U(\{\mathbf{r}_i\}) = U_{\text{DD}} - E_{\text{spin}}^{\text{ground}},
\end{equation}
where $U_{\text{DD}}$ is the ``magnetic'' embedded atom potential developed by Dudarev and Derlet~\cite{Dudarev2005, Derlet2007},
and ${E_{\text{spin}}^{\text{ground}} = - \sum_{i<j} J_{ij}(\left\{ \mathbf{r}_k \right\})|\mathbf{S}_i||\mathbf{S}_j|}$ 
is the energy contribution from a collinear spin state,
subtracted out to eliminate the magnetic interaction energy that is implicitly contained in $U_{\text{DD}}$.
With the particular form of $U(\{\mathbf{r}_i\})$ given in Eq.~\eqref{eq:atomic_potential},
Hamiltonian~\eqref{eq:hamiltonian} provides the same ground state energy as $U_{\text{DD}}$.
The exchange interaction is modeled via a simple pairwise function $J(r_{ij})$ parameterized by first-principles calculations~\cite{Ma2008},
with spin lengths absorbed into its definition, i.e. ${J(r_{ij}) = J_{ij}(\left\{ \mathbf{r}_k \right\}) |\mathbf{S}_i||\mathbf{S}_j|}$.
We assume constant spin lengths $|\mathbf{S}| = 2.2/g$, with $g$ being the electron $g$ factor. 

We would like to point out that the fluctuation of the magnitudes of magnetic moments  
and spin-orbit interactions are not considered in this work.
In transition metals and alloys, fluctuation of spin magnitudes may have a notable effect on the material properties, particularly at high temperatures. 
Ma \textit{et al.}~\cite{LongFluc1} proposed a way of incorporating longitudinal spin fluctuations into SD and MD-SD simulations via a Langevin-type equation of motion
within the context of fluctuation-dissipation theorem.
Numerical coefficients of the corresponding Landau Hamiltonian can be determined from \textit{ab initio} calculations~\cite{LongFluc1, LongFluc2}.
An accurate depiction of spin-orbit interactions can be potentially achieved with the use of Hubbard-like Hamiltonians 
as the foundation for deriving the equations of motion~\cite{spin_orbit_2}.
A phenomenological approach for modeling spin-orbit interactions in MD-SD  has also been recently proposed~\cite{spin_orbit}, 
but was not adopted in this study due to its computationally demanding nature.

\subsection{Characterizing collective excitations}
In MD and SD simulations, space-displaced, time-displaced correlation functions of the microscopic dynamical variables are
integral to the study of the collective phenomena in the system~\cite{Rapaport2004, Lovesey1984, Landau1999}.
Fourier transforms of these quantities directly yield information regarding the frequency spectra and the lifetimes of the 
respective collective modes.

Let us define microscopic atom density as
\begin{equation}
	\rho_n (\mathbf{r}, t) = \sum_i \delta \left[\mathbf{r}-\mathbf{r}_i(t)\right].
\end{equation}
The spatial Fourier transform of the space-displaced, time-displaced density-density correlation function, 
namely, the intermediate scattering function~\cite{Hansen2006} then takes the form
\begin{equation} \label{eq:Fnn}
	F_{nn}(\mathbf{q}, t) = \frac{1}{N} \left< \rho_n(\mathbf{q}, t) \rho_n(-\mathbf{q}, 0)\right >,
\end{equation}
where $ \rho_n(\mathbf{q}, t) = \int \rho_n(\mathbf{r}, t) e^{ -i\mathbf{q} \cdot \mathbf{r} } d\mathbf{r}  = \sum_i e^{ -i\mathbf{q} \cdot \mathbf{r}_i(t) }$.
The power spectrum of the intermediate scattering function
\begin{equation} \label{eq:Snn}
	S_{nn}(\mathbf{q}, \omega) = \frac{1}{2\pi} \int_{-\infty}^{+\infty} F_{nn}(\mathbf{q}, t) e^{-i\omega t} dt,
\end{equation}
is called the ``density-density dynamic structure factor'' for the momentum transfer $\mathbf{q}$ and frequency (energy) transfer $\omega$.
$S_{nn}(\mathbf{q}, \omega)$ is directly related to the differential cross section measured in inelastic neutron scattering experiments~\cite{Hansen2006}.
Local density fluctuations in a system are caused by the thermal diffusion of atoms as well 
as vibrational modes related to the propagating lattice waves~\cite{Anento2004}.
For liquid systems, the thermal diffusive mode can be identified as a peak in $S_{nn}(\mathbf{q}, \omega)$ centered at $\omega = 0$,
whereas for solids this peak will disappear due to the absence of thermal diffusion~\cite{Anento2004}.
In crystalline solids, peaks in $S_{nn}(\mathbf{q}, \omega)$ at non-zero frequencies can be uniquely associated with 
longitudinal vibrational modes with the corresponding frequencies and wave vectors.
As the transverse lattice vibrations do not cause local density fluctuations towards the direction of wave propagation,
$S_{nn}(\mathbf{q}, \omega)$ is incapable of revealing 
information about these modes.
Therefore, to identify transverse lattice vibrations, one needs to consider the time-dependent correlations of transverse velocity components.

With the microscopic ``velocity density'' defined as $\boldsymbol{\rho}_v(\mathbf{r}, t) = \sum_i \mathbf{v}_i(t) \delta \left[\mathbf{r}-\mathbf{r}_i(t)\right]$,
the spatial Fourier transform of the  velocity-velocity correlation function takes the form
\begin{equation} \label{eq:Fvv}
	F_{vv}^{L,T}(\mathbf{q}, t) = \frac{1}{N} \left< \boldsymbol{\rho}_v^{L,T}(\mathbf{q}, t) \cdot \boldsymbol{\rho}_v^{L,T}(-\mathbf{q}, 0)\right >,
\end{equation}
where $\boldsymbol{\rho}_v^{L,T}(\mathbf{q}, t) = \sum_i \mathbf{v}_i^{L,T}(t)  e^{ -i\mathbf{q} \cdot \mathbf{r}_i(t) }$,
with the superscripts $L$ and $T$ respectively denoting the longitudinal and transverse components with reference to the direction of the wave propagation.
Peaks in the corresponding power spectra $S_{vv}^L(\mathbf{q}, \omega)$ and $S_{vv}^T(\mathbf{q}, \omega)$ 
respectively reveal longitudinal and transverse vibrational modes of the system.
It can be shown that $S_{vv}^L(\mathbf{q}, \omega)$ is directly related to the density-density dynamic structure factor $S_{nn}(\mathbf{q}, \omega)$ 
via the relationship $S_{nn}(\mathbf{q}, \omega) = \omega^2/q^2 S_{vv}^L(\mathbf{q}, \omega)$~\cite{Hansen2006, Anento2004}.

Just as the time-dependent density-density and velocity-velocity correlations reveal vibrational excitations associated with the lattice subsystem,
spin density autocorrelations can elucidate the magnetic excitations associated with the spin subsystem.

The microscopic ``spin density'' is given by
\begin{equation}
	\boldsymbol{\rho}_s (\mathbf{r}, t) = \sum_i \mathbf{S}_i(t) \delta \left(\mathbf{r}-\mathbf{r}_i(t)\right).
\end{equation}
Treating the spin-spin correlations along $x$, $y$, and $z$ directions separately, 
we define the intermediate scattering function as
\begin{equation} \label{eq:Fss}
	F_{ss}^k(\mathbf{q}, t) = \frac{1}{N} \left< \rho_s^k(\mathbf{q}, t) \rho_s^k(-\mathbf{q}, 0)\right >,
\end{equation} 
where $k = x, y,$ or $z$, and 
$ \boldsymbol{\rho}_s(\mathbf{q}, t) = \sum_i \mathbf{S}_i(t) e^{-i\mathbf{q} \cdot \mathbf{r}_i(t)} $.
For a ferromagnetic system in the microcanonical ensemble, the magnetization vector is a constant of motion
and serves as a fixed symmetry axis throughout the time evolution of the system.
To differentiate between the magnetic excitations that propagate parallel and perpendicular to this symmetry axis,
we redefine the coordinate system in spin space such that the $z$ axis is parallel to the magnetization vector.
The components $\{F_{ss}^k(\mathbf{q}, t)\}$  can then be simply regrouped to yield the longitudinal component
\begin{equation}
	F_{ss}^L(\mathbf{q}, t) = F_{ss}^z(\mathbf{q}, t),
\end{equation}
and the transverse component 
\begin{equation}
	F_{ss}^T(\mathbf{q}, t) = \frac{1}{2} \left( F_{ss}^x(\mathbf{q}, t) + F_{ss}^y(\mathbf{q}, t) \right).
\end{equation}
Note that the separation of magnetic excitations into longitudinal and transverse modes is only meaningful for temperatures below the Curie temperature $T_\text{C}$,
since above $T_\text{C}$, the net magnetization vanishes and all directions in spin space become equivalent.

Fourier transforms of $F_{ss}^{L,T}(\mathbf{q}, t)$ yield the spin-spin dynamic structure factors $S_{ss}^{L,T}(\mathbf{q}, \omega)$.
Just like the density-density dynamic structure factor, the spin-spin dynamic structure factor
is a measurable quantity in inelastic neutron scattering experiments~\cite{Lovesey1984, Landau1999}.

In this study, we are primarily interested in investigating wave propagation along the three principle lattice directions:
[100], [110] and [111]. Let us denote the wave vectors in these directions as $\mathbf{q} = (q, 0, 0)$, $(q, q, 0)$, and $(q, q, q)$, respectively.
Due to the finite size of the simulation box, the accessible values of $q$ in each direction is constrained to a discrete set given by
$q = 2 \pi n_q/La$, with $n_q = \pm 1, \pm 2, \dotsc, \pm,  L$ for the [100] and [111] directions, and $n_q = \pm 1, \pm 2, \dotsc, \pm,  L/2$ for the [110] direction,
where $L$ is the linear lattice dimension and $a = 2.8665$\;\AA~is the lattice constant of bcc iron.

\subsection{Simulation details} \label{sec:sim_details}

For integrating the coupled equations of motion presented in Eq.~\eqref{eq:eom}, we adopted an algorithm based on 
the second order Suzuki-Trotter (ST) decomposition of the non-commuting operators~\cite{Omelyan2001, Tsai2005, Krech1998}.
To obtain a reasonable level of accuracy as reflected by the energy and magnetization conservation,
an integration time step of $\Delta t = 1$\,fs was used. 

For computing canonical averages of time-dependent correlation functions, we used time series obtained from microcanonial dynamical simulations,
that are, in turn, initiated from equilibrium states drawn from the canonical ensemble at the desired temperature $T$.
Averaging over the results of multiple simulations started from different initial states 
yields good estimates of the respective canonical ensemble averages~\cite{Landau1999}.

\begin{figure}
 \includegraphics[width=\columnwidth]{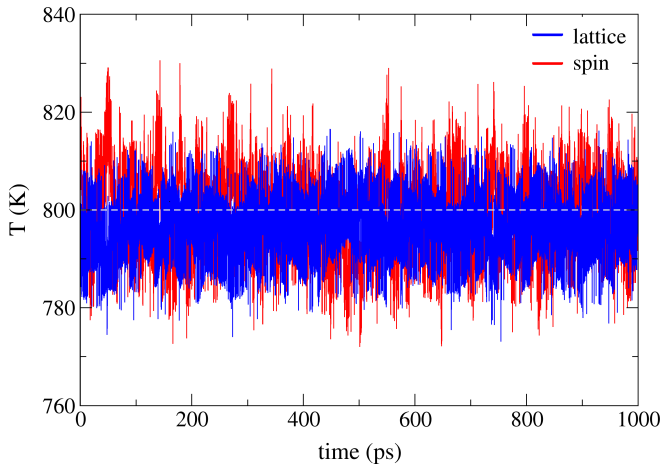}
  \caption{ 
	  Time evolution of the instantaneous lattice and spin temperatures 
	  as observed in a microcanonical MD-SD simulation for the system size $L=16$ at temperature $T=800$\;K.
	  The initial state for the time integration was generated from the procedure described in Sec.~\ref{sec:sim_details}.
	  The spin temperature was measured using the formula developed by Nurdin \textit{et al.}~ \cite{Nurdin2000}.
	  }	
    \label{fig:T_vs_t.eps}
\end{figure}

For generating the initial states for our microcanonical MD-SD simulations, we adhere to the following procedure.
First, we equilibrate the subspace consisting of positions and spins using the Metropolis Monte Carlo (MC) method~\cite{Metropolis1953}. 
As the second step, we assign initial velocities to the atoms based on the Maxwell-Boltzmann distribution at the desired temperature $T$.
Finally, we perform a short microcanonical MD-SD equilibration run (typically $\sim 1000$ time steps with $\Delta t = 1$\;fs), 
which would ultimately bring the whole system to the equilibrium by resolving any inconsistencies 
between the position-spin subspace and the velocity distribution.
Fig.~\ref{fig:T_vs_t.eps} shows the time evolution of the instantaneous lattice and spin temperatures 
as observed in a microcanonical MD-SD simulation initiated from an equilibrium state generated from the aforementioned technique for $T=800$\;K.
Both lattice and spin temperatures fluctuate about a mean value of $T=800$\;K, 
indicating that the lattice and the spin subsystems are in mutual equilibrium.

To characterize phonon and magnon modes, 
we performed simulations for the system size $L=16$ ($8192$ atoms) at temperatures $T=300$\,K, $800$\,K, and $1000$\,K.
$T = 1000$\,K was particularly chosen due to its vicinity to the Curie temperature of bcc iron, $T_C \approx 1043$\,K.
(A recent high resolution Monte Carlo study has revealed that the transition temperature of the particular spin-lattice model
used in our study to be $T \approx 1078$\,K~\cite{rewl}). 
Equations of motion were integrated up to a total time of $t_\text{max} = 1$\,ns, and the space-displaced, time-displaced correlation functions were computed 
for the three principle lattice directions: [100], [110] and [111].
To increase the accuracy, we have averaged these quantities over different starting points in the time series.
Canonical ensemble averages were estimated using the results of $200$ independent simulations, each initiated from a different initial state.
The time Fourier transform in Eq.~\eqref{eq:Snn} was carried out to a cutoff time of $t_\text{cutoff} = 0.5$\,ns.

As our primary goal is to understand the mutual impact of the phonons and magnons on their respective frequency spectra and lifetimes,
we have also performed standalone MD and SD simulations for comparison. 
For MD simulations, we used the Dudarev-Derlet potential to model the interatomic interactions
while completely neglecting the spin-spin interactions.
SD simulations were conducted with the atoms frozen at perfect bcc lattice positions, and the exchange parameters
determined from the same pairwise function used for MD-SD simulations.

\section{Results} \label{sec:results}

\subsection{Vibrational excitations} \label{sec:results_1}

\begin{figure}
 \includegraphics[width=\columnwidth]{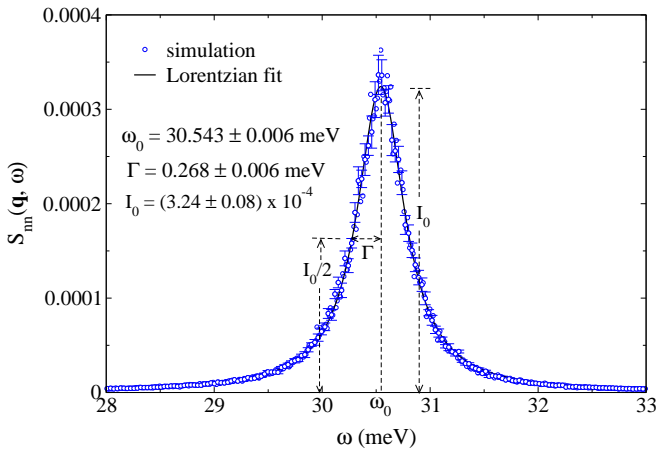}
  \caption{ 
	  Density-density dynamic structure factor for $\mathbf{q} = (1.1$\,\AA\textsuperscript{-1}, $0, 0)$
	  obtained from MD-SD simulations for $L=16$ at $T=300$\;K.
	  The symbols represent simulation data while the solid line is a fit with the Lorentzian lineshape given in Eq.~\eqref{eq:lorentz_fit}.
	  }	
    \label{fig:Snn_fitting}
\end{figure}

For all the temperatures considered, we observe well defined excitation peaks at non-zero frequencies in the 
density-density dynamic structure factor $S_{nn}(\mathbf{q}, \omega)$, as well as in 
the longitudinal and the transverse components of the velocity-velocity dynamic structure factor: $S_{vv}^L(\mathbf{q}, \omega)$ and $S_{vv}^T(\mathbf{q}, \omega)$.
For each $\mathbf{q}$ along [100] and [111] lattice directions, all three quantities show single peaks (See Fig.~\ref{fig:Snn_fitting} for an example).
The peak positions in $S_{nn}(\mathbf{q}, \omega)$ and $S_{vv}^L(\mathbf{q}, \omega)$ for the same wave vector coincide with each other as they 
are both associated with the longitudinal vibrational modes, and hence convey the same information.
The peaks in $S_{vv}^T(\mathbf{q}, \omega)$ are associated with the transverse lattice vibrations.
Since there are two orthogonal directions perpendicular to a given wave vector $\mathbf{q}$,
there are, in fact, two transverse vibrational modes for each $\mathbf{q}$.
Due to the four-fold and three-fold rotational symmetry about the axes [100] and [111], respectively,
the two transverse modes for the wave vectors along these directions become degenerate~\cite{LatticeDynamics}. 
As a result, we only observe a single peak in $S_{vv}^T(\mathbf{q}, \omega)$ for the wave vectors along these directions.

We also observe single peak structures in $S_{nn}(\mathbf{q}, \omega)$ and $S_{vv}^L(\mathbf{q}, \omega)$ for the wave vectors along the [110] direction.
However, for the case of $S_{vv}^T(\mathbf{q}, \omega)$, one can clearly identify two distinct peaks. 
This is a consequence of the two transverse modes being non-degenerate due to the reduced rotational symmetry (two-fold) about the [110] axis
in comparison to [100] and [111] directions~\cite{LatticeDynamics}.

To extract the positions and the half-widths of the phonon peaks, 
we fit the simulation results for the dynamic structure factor to a Lorentzian function of the form~\cite{Tsai2000, Tao2005}
\begin{equation} \label{eq:lorentz_fit}
	S(\mathbf{q}, \omega) = \frac{I_0 \Gamma^2}{(\omega-\omega_0)^2 + \Gamma^2},
\end{equation}
where $\omega_0$ is the characteristic frequency of the vibrational mode, $I_0$ is the intensity or the amplitude of the peak,
and $\Gamma$ is the half-width at half maximum (HWHM) which is inversely proportional to the lifetime of the excitation.
The errors of the fitting parameters were estimated using the following procedure.
The complete set of correlation function estimates obtained from $200$ independent simulations was divided into $10$ groups, 
and the data within each group were averaged over to yield $10$ results sets. 
Dynamic structure factors were independently computed for these $10$ correlation function sets. 
To estimate the errors in the fitting parameters, 
we separately performed curve fits to these $10$ independent dynamic structure factor estimates, 
and calculated the standard deviations of the fitting parameters.
Statistical errors bars obtained in this manner were found to be an order of magnitude larger than the error bars estimated by the curve-fitting tool.

For all the temperatures and wave vectors considered, 
the Lorentzian lineshape given in Eq.~\eqref{eq:lorentz_fit} fitted well with the peaks observed in $S_{nn}(\mathbf{q}, \omega)$ and $S_{vv}^{L,T}(\mathbf{q}, \omega)$.  
Fig.~\ref{fig:Snn_fitting} shows an example curve fit for the MD-SD results of $S_{nn}(\mathbf{q}, \omega)$ 
for $\mathbf{q} = (1.1$\,\AA\textsuperscript{-1},$0, 0)$ at $T=300$\,K.
To fit the two peak structure observed in $S_{vv}^T(\mathbf{q}, \omega)$ for the [110] direction, 
we use the sum of two Lorentzians.

\begin{figure}
 \includegraphics[width=\columnwidth]{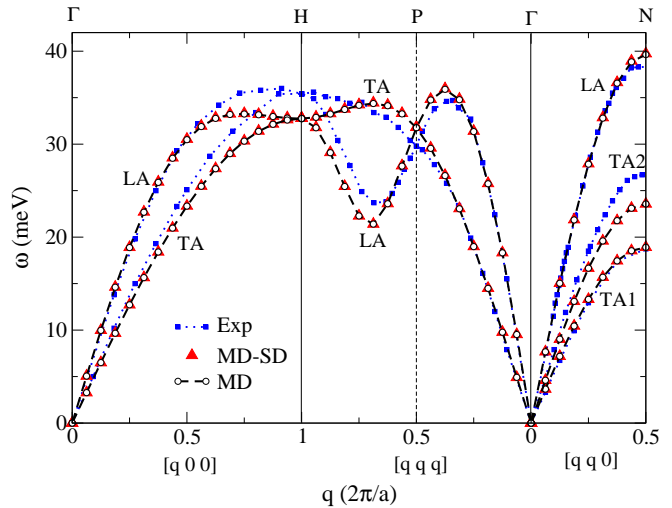}
  \caption{ 
	  Comparison of the phonon dispersion curves obtained from MD-SD simulations ($L=16$) with the experimental results~\cite{phononExp1, phononExp2}
	  for $T = 300$\;K.
	  Results obtained from pure MD simulations are also plotted for comparison.
	  LA and TA, respectively, denote the longitudinal and transverse branches.
	  }	
    \label{fig:phonon_dispersion_T_300}
\end{figure}

Using the peak positions obtained from the Lorentzian fits, one can construct phonon dispersion relations for the three principle lattice directions.
Fig.~\ref{fig:phonon_dispersion_T_300} shows the the dispersion curves determined from our MD-SD simulations for $T=300$\,K, 
along with the experimental results~\cite{phononExp1, phononExp2} obtained from inelastic neutron scattering.  
For comparison, we have also shown the results of standalone MD simulations for the same temperature.
In general, for small to moderate $q$ values, both MD-SD and MD dispersion curves agree well with the experimental results,
but deviations can be observed for larger $q$ values, particularly near the zone boundaries in [100] and [111] directions. 
Although the MD-SD and MD dispersion curves are indistinguishable within the resolution of Fig.~\ref{fig:phonon_dispersion_T_300},
we will show later on that there are, in fact, deviations larger than the error bars. 

At temperatures in the vicinity of absolute zero, 
due to the low occupation of vibrational modes,
phonons behave as weakly interacting quasiparticles that can be treated within the harmonic approximation~\cite{phononScattering1}.
In this limit, characteristic frequencies of the phonons are well defined
and the lifetimes are practically infinite.
As the temperature is increased, phonon occupation numbers also increase,
which in turn increases the probability of mutual interactions.
As a result of such phonon-phonon scattering at elevated temperatures,
characteristic frequencies of the phonons may shift,
and the lifetimes may shorten~\cite{phononScattering1, phononScattering2}.
In magnetic crystals, the co-existence of phonons and magnons gives rise to another class of scattering processes,
namely, phonon-magnon scattering.
Just as phonon-phonon scattering, phonon-magnon scattering 
may also lead to a shift in the characteristic phonon frequencies,
as well as shortening of the phonon lifetimes.
As the occupancy of both phonon and magnon modes increases with temperature, 
these effects will be more pronounced as the temperature is increased.

\begin{figure}
 \includegraphics[width=\columnwidth]{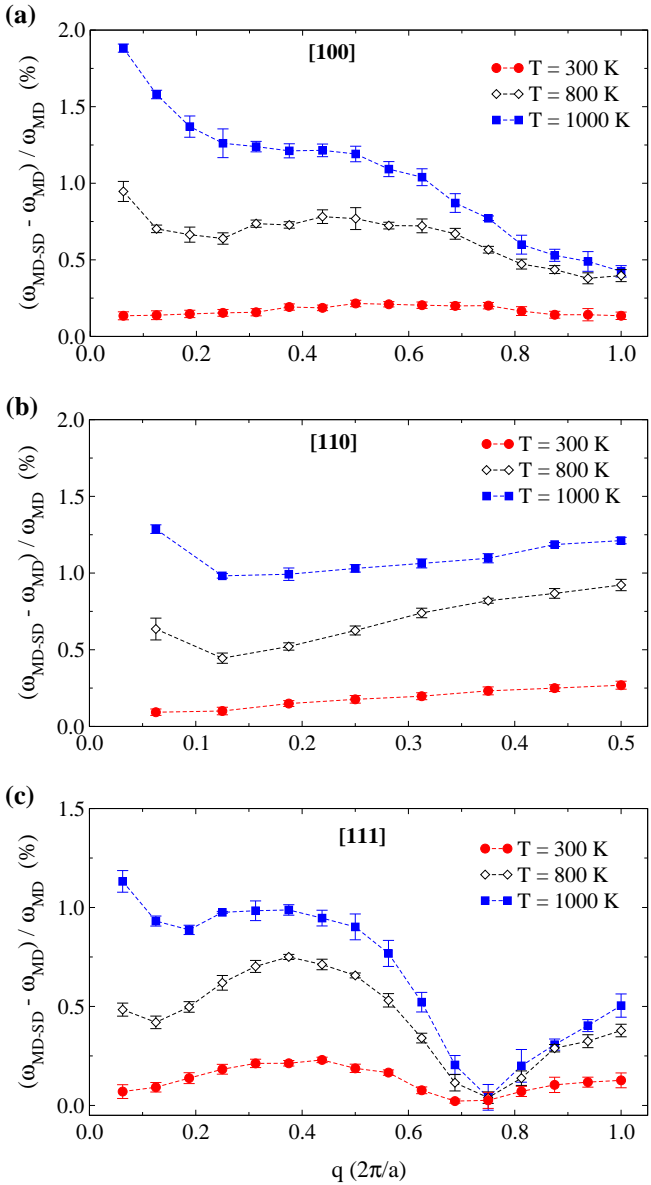}
  \caption{ 
	  The fractional shift in longitudinal phonon frequencies due to magnons for $L=16$ at $T=300$\;K, $T=800$\;K, and $T=1000$\;K
	  in the (a) [1 0 0], (b) [1 1 0], and (c) [1 1 1] lattice directions.
	  }	
    \label{fig:long_phonon_freq_shift}
\end{figure}

\begin{figure}
 \includegraphics[width=\columnwidth]{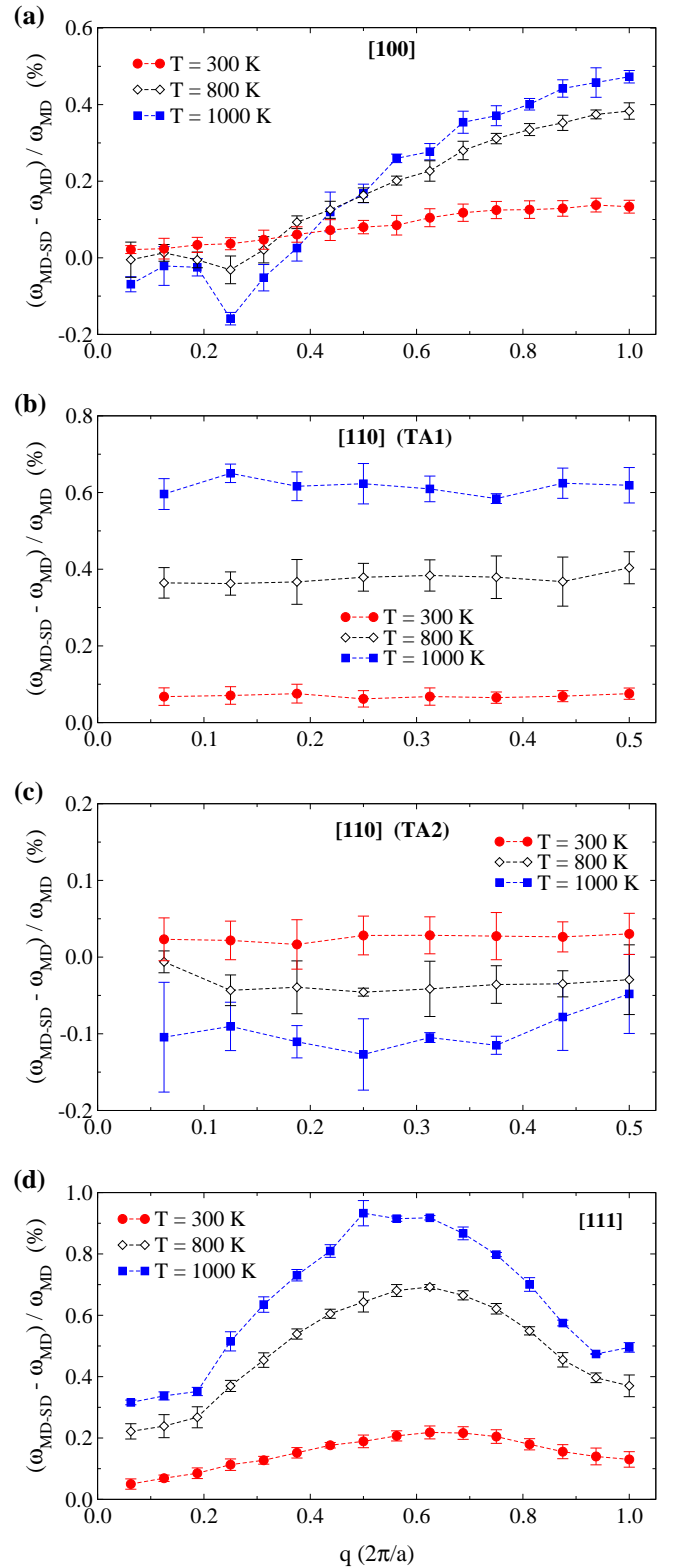}
  \caption{ 
	  The fractional shift in transverse phonon frequencies due to magnons for $L=16$ at $T=300$\;K, $T=800$\;K, and $T=1000$\;K
	  in the (a) [1 0 0], (b) [1 1 0] (TA1), (c) [1 1 0] (TA2), and (d) [1 1 1] lattice directions.
	  }	
    \label{fig:trans_phonon_freq_shift}
\end{figure}

To carefully examine the changes in the phonon frequency spectrum due to magnons,
we compare the characteristic frequencies determined from MD-SD simulations ($\omega_{_\text{MD-SD}})$ with the ones obtained from 
MD simulations ($\omega_{_\text{MD}}$) by calculating the fractional frequency shift, 
${ \left( \omega_{_\text{MD-SD}} - \omega_{_\text{MD}} \right)/\omega_{_\text{MD}} }$.
The results for the three principle directions are shown in Figs.~\ref{fig:long_phonon_freq_shift} and~\ref{fig:trans_phonon_freq_shift},
for the longitudinal and the transverse modes, respectively.
With the exception of the high frequency transverse branch along [110] direction (TA2), 
phonon frequencies shift to higher values in the presence of magnons.
In general, the shift in frequencies becomes more pronounced as the temperature is increased.
A particularly interesting behavior occurs in the longitudinal branch for the [111] direction
where we observe dips in the curves for all three temperatures at the same $q$ value. 
For all three temperatures, the frequency shift of the vibrational mode that corresponds to the bottom of the dip is close to zero.
Therefore, the frequency of this phonon mode appears to be unaffected by the presence of magnons.

\begin{figure}
 \centering
 \includegraphics[width=\columnwidth]{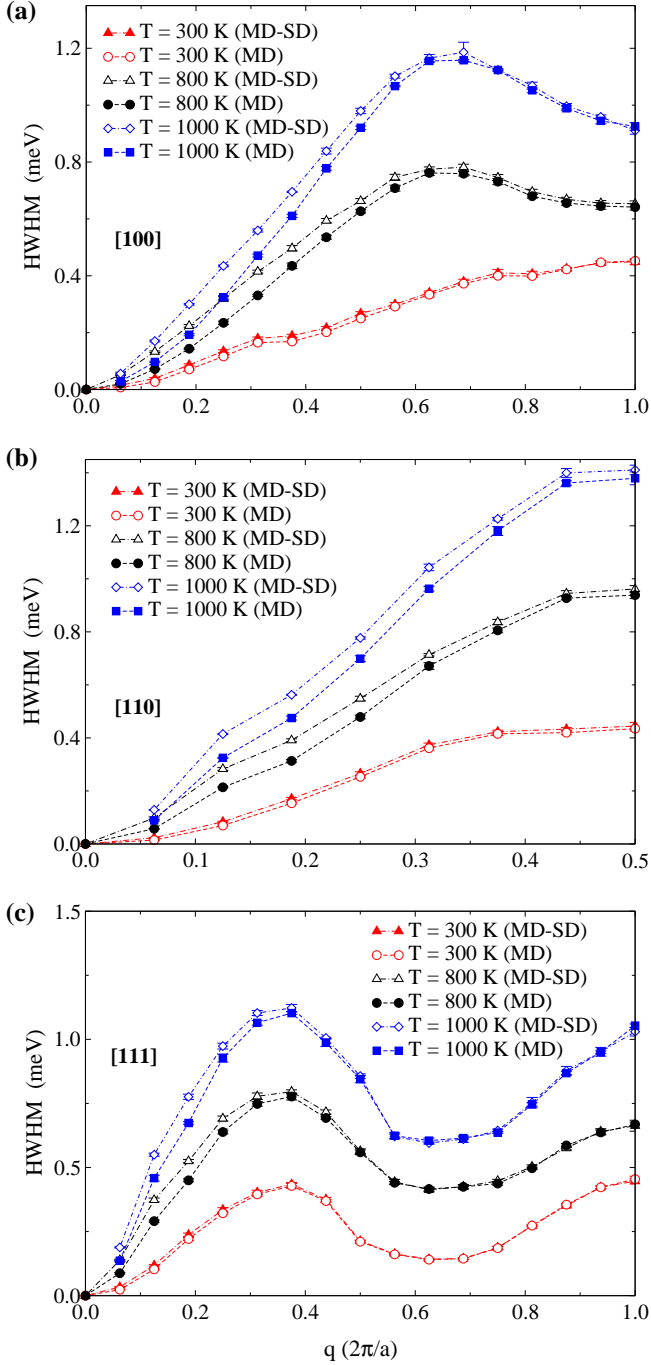}
  \caption{ 
	  Half-width at half maximum (HWHM) of the longitudinal phonons at $T=300$\;K, $T=800$\;K, and $T=1000$\;K
	  obtained from MD-SD and MD simulations for $L=16$
	  in the (a) [1 0 0], (b) [1 1 0], and (c) [1 1 1] lattice directions.
	  }	
    \label{fig:long_phonon_halfwidth}
\end{figure}

Lifetimes of the phonon excitations are inversely proportional to the half-widths at half maximum of the corresponding vibrational peaks
observed in $S_{nn}(\mathbf{q}, \omega)$ and $S_{vv}^{L,T}(\mathbf{q}, \omega)$.
To study the impact of the magnons on the phonon lifetimes, we compare the half-widths 
obtained from MD-SD simulations with that of the MD simulations.
Fig.~\ref{fig:long_phonon_halfwidth} shows the results for the longitudinal phonons.
For the longitudinal phonons at $T=300$\,K, a marginal increase in the half-widths can be observed due to the magnons,
which becomes more pronounced as the temperature is increased.
For the case of transverse phonons, we did not observe any notable difference between the MD-SD and MD half-widths outside the error bars, 
for all the temperatures considered.

\subsection{Magnetic excitations} \label{sec:results_2}

\subsubsection{Transverse magnon modes}

\begin{figure}
 \includegraphics[width=\columnwidth]{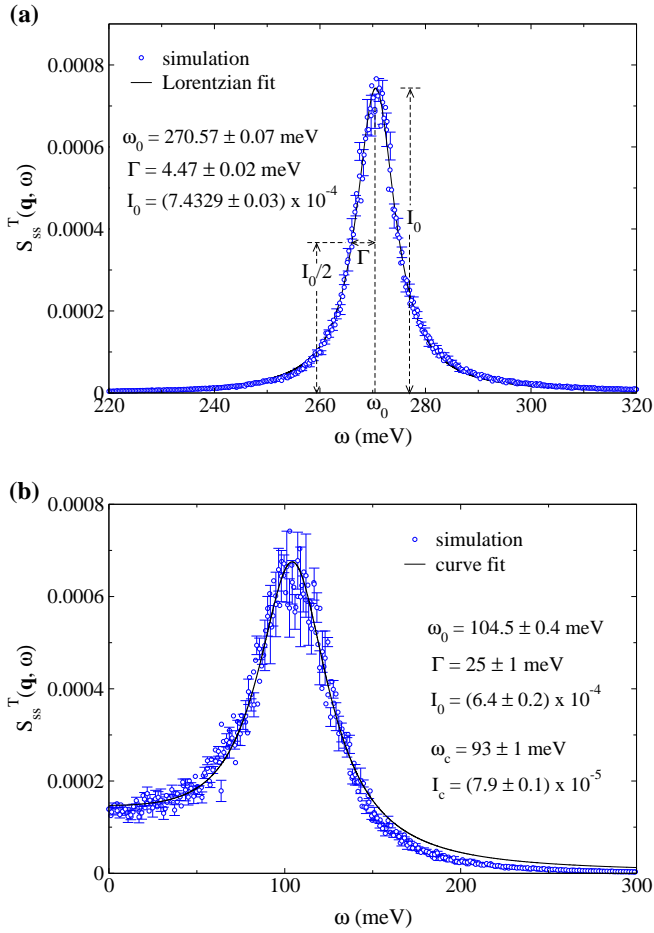}
  \caption{ 
	  Transverse spin-spin dynamic structure factor obtained from MD-SD simulations for $L=16$.
	  (a) $T=300$\;K and $\mathbf{q} = (1.1$\,\AA\textsuperscript{-1}, $0, 0)$, (b) $T=1000$\;K and $\mathbf{q} = (0.82$\,\AA\textsuperscript{-1}, $0, 0)$.
	  The symbols represent simulation data while the solid lines are fits to functional forms presented by
	  Eq.~\eqref{eq:lorentz_fit} [for panel (a)] and Eq.~\eqref{eq:lorentz_gauss_fit} [for panel (b)].
	  }
    \label{fig:Sss_T_fitting}
\end{figure}

For the temperatures $T=300$\,K and $T=800$\,K, our results for the transverse component of the spin-spin dynamic structure factor $S_{ss}^T(\mathbf{q}, \omega)$
show a single spin wave peak, that can be fitted to a Lorentzian lineshape of the form Eq.~\eqref{eq:lorentz_fit} (See Fig.~\ref{fig:Sss_T_fitting} (a) for an example.).
For $T=1000$\,K, we also observe a diffusive central peak at $\omega = 0$, as observed in neutron scattering experiments~\cite{central_peak} 
and previous SD studies~\cite{Tao2005}.
This two-peak structure can be best captured by a function of the form~\cite{Tao2005}
\begin{equation} \label{eq:lorentz_gauss_fit}
	S(\mathbf{q}, \omega) = I_c \exp\left(-\omega^2/\omega_c^2\right) + \frac{I_0 \Gamma^2}{(\omega-\omega_0)^2 + \Gamma^2},
\end{equation}
where the first term (Gaussian) corresponds to the central peak, and the second term (Lorentzian) describes the spin wave peak
(See Fig.~\ref{fig:Sss_T_fitting} (b) for an example.).
For large $\mathbf{q}$ values at $T=800$\,K and $T=1000$\,K, spin wave peaks in $S_{ss}^T(\mathbf{q}, \omega)$  were found to be asymmetric, 
and hence did not yield good fits to Lorentzian lineshapes.
Therefore, one cannot obtain reliable estimates of the magnon half-widths.
However, spin wave peak positions can still be determined relatively precisely, thus the magnon dispersion relations can be constructed.

\begin{figure}
 \includegraphics[width=\columnwidth]{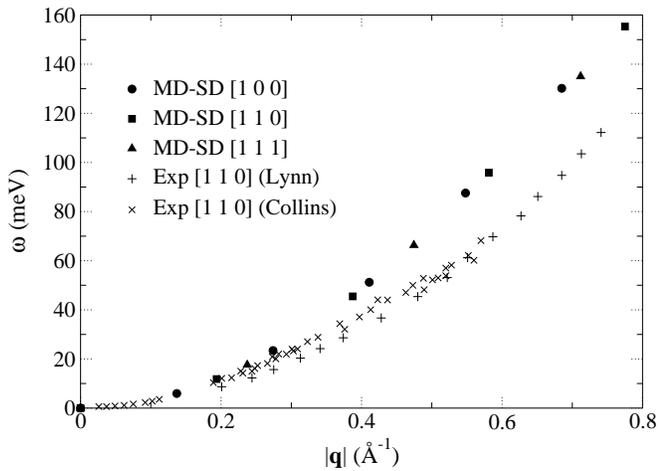}
 \caption{ 
	  Transverse magnon dispersion relations at $T = 300$\;K obtained from MD-SD simulations for $L=16$.
	  The experimental results reported by Lynn~\cite{Lynn} and Collins~\cite{Collins} for the [1 1 0] direction 
	  are also plotted for comparison.
	  }	
    \label{fig:magnon_dispersion_small_q}
\end{figure}

\begin{figure}
 \includegraphics[width=\columnwidth]{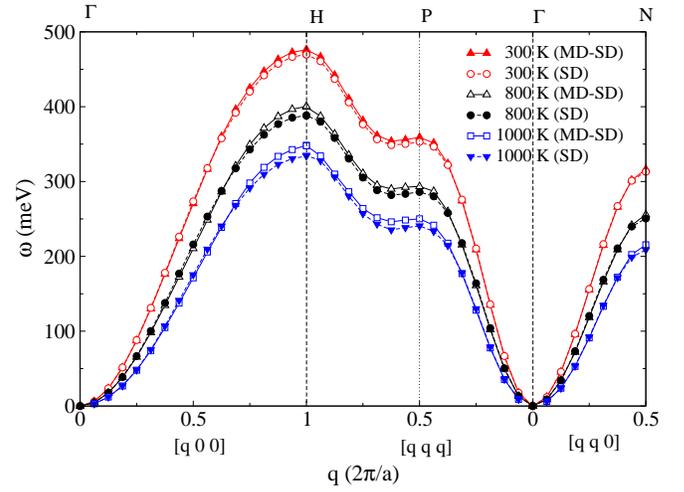}
  \caption{ 
	  Transverse magnon dispersion curves for $T = 300$\;K, $T = 800$\;K, and $T = 1000$\;K obtained from MD-SD and SD simulations for $L=16$.
	  }	
    \label{fig:magnon_dispersion}
\end{figure}

Fig.~\ref{fig:magnon_dispersion_small_q} shows the transverse magnon dispersion relations for small $|\mathbf{q}|$ values along the three principle directions 
as determined from MD-SD simulations at $T=300$\,K.
In agreement with the experimental findings~\cite{Lynn, Collins}, the three dispersion relations are isotropic when plotted as
functions of the magnitude of the wave vector $|\mathbf{q}|$.
Moreover, for small $|\mathbf{q}|$ values, our results agree quantitatively with the experimental results for the [110] direction~\cite{Lynn, Collins}.
Fig.~\ref{fig:magnon_dispersion} shows the complete dispersion curves determined from MD-SD and SD simulations for $T=300$\,K, $T=800$\,K, and $T=1000$\,K.
For both MD-SD and SD, the characteristic frequencies shift to lower values as the temperature is increased.
This indicates increased magnon-magnon scattering at elevated temperatures.
For all three temperatures, particularly near the zone boundaries, we can observe a marginal difference between the MD-SD and SD dispersion curves.
This, in fact, is a result of phonon-magnon scattering. 

\begin{figure}
 \includegraphics[width=\columnwidth]{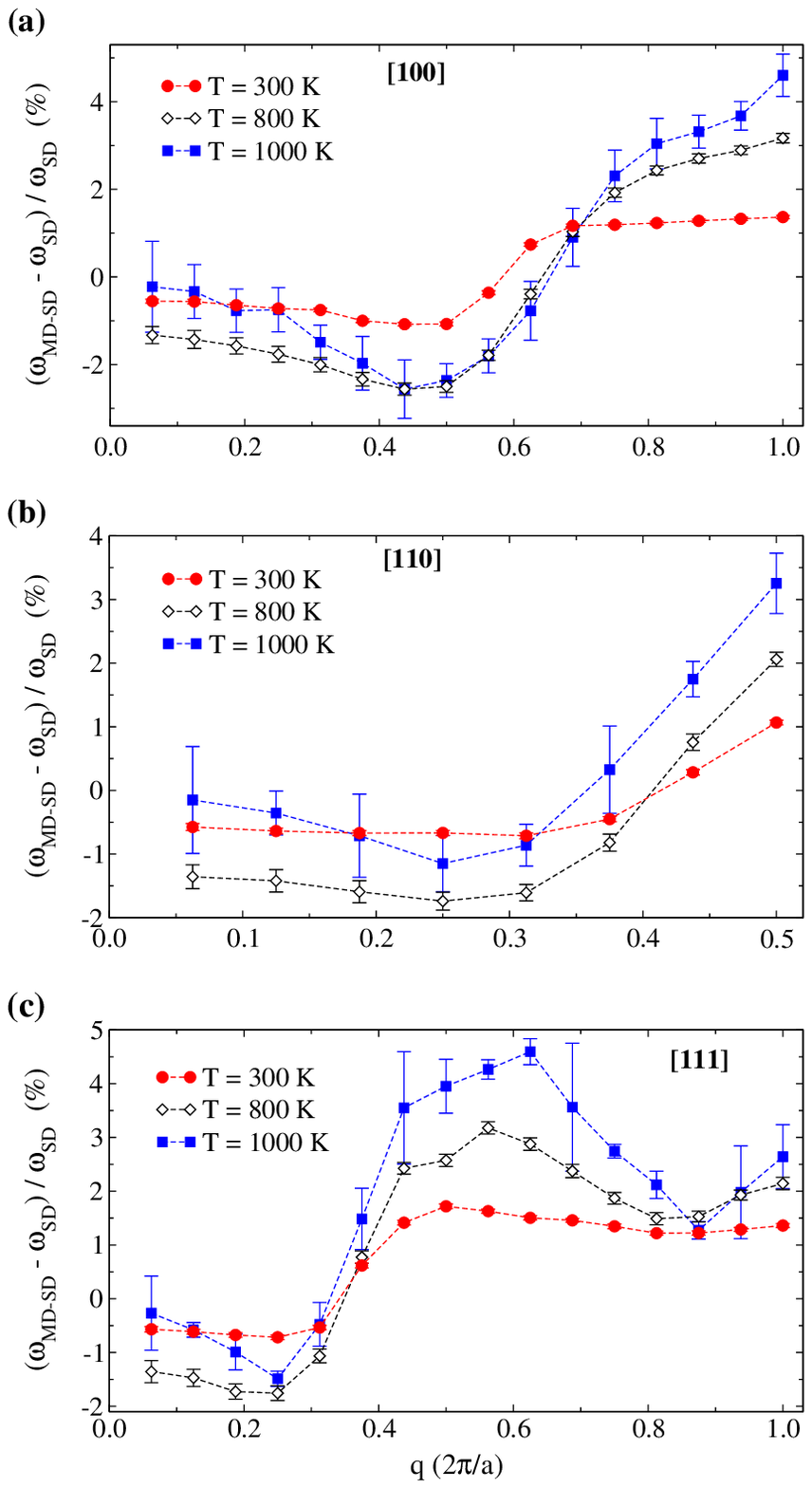}
  \caption{ 
	  The fractional shift in transverse magnon frequencies due to phonons for $L=16$ at $T=300$\;K, $T=800$\;K, and $T=1000$\;K
	  in the (a) [1 0 0], (b) [1 1 0], and (c) [1 1 1] lattice directions.
	  }	
    \label{fig:magnon_freq_shift}
\end{figure}

\begin{figure}
 \includegraphics[width=\columnwidth]{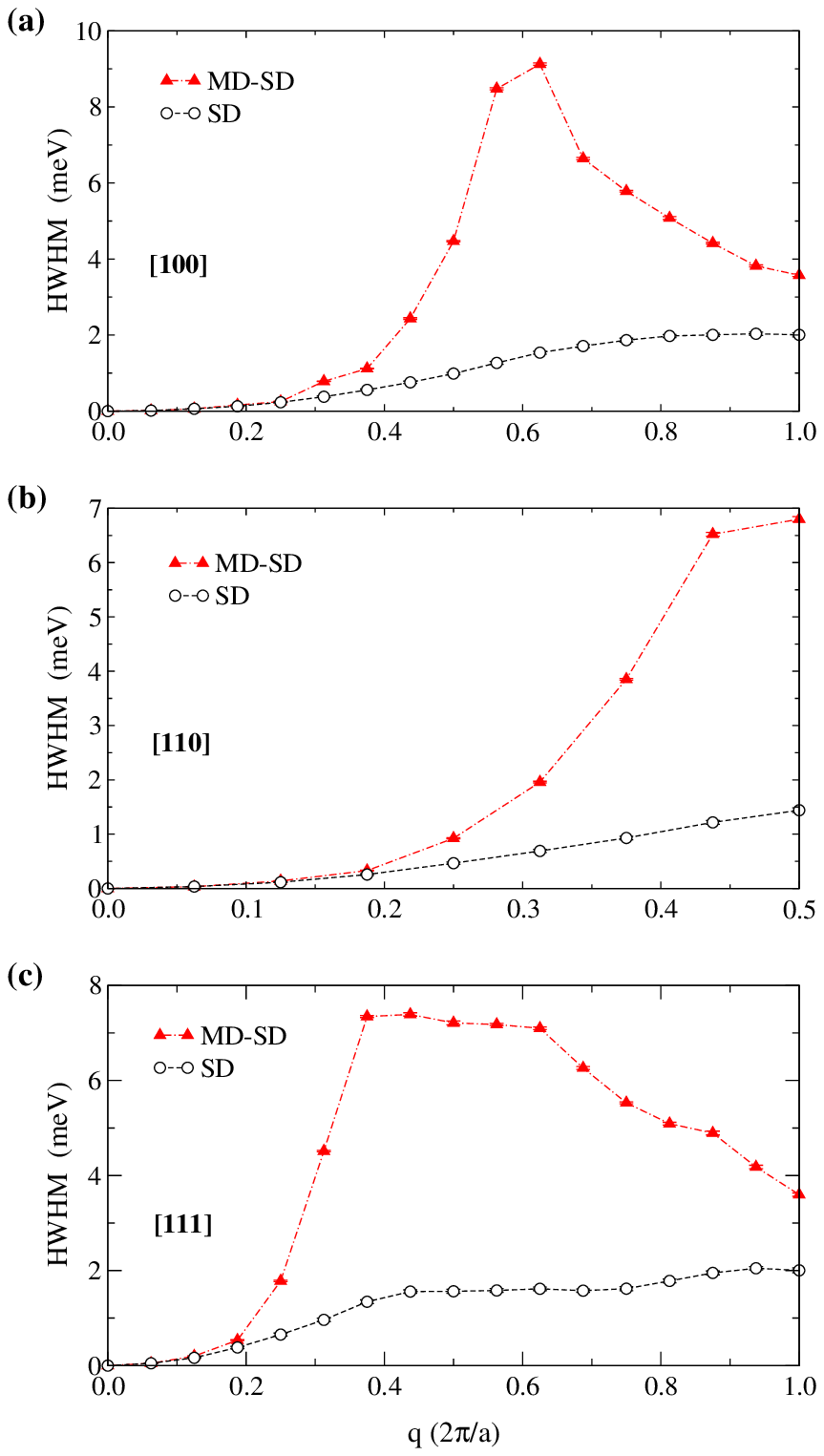}
  \caption{ 
	  Half-width at half maximum (HWHM) of the transverse magnons at $T=300$\;K
	  obtained from MD-SD and SD simulations for $L=16$
	  in the (a) [1 0 0], (b) [1 1 0], and (c) [1 1 1] lattice directions.
	  }	
    \label{fig:magnon_halfwidth}
\end{figure}

To further investigate the magnon softening due to phonons, 
we calculate the fractional frequency shift of the magnons, 
${ \left( \omega_{_\text{MD-SD}} - \omega_{_\text{SD}} \right)/\omega_{_\text{SD}} }$.
The results are shown in Fig.~\ref{fig:magnon_freq_shift} for the three principle directions.
For small $q$ values, magnon modes shift to lower frequencies in the presence of phonons.
As $q$ increases, the direction of the shift is reversed.
Moreover, the shift in frequencies becomes more pronounced as the temperature is increased.

Fig.~\ref{fig:magnon_halfwidth} compares the transverse magnon half-widths obtained from MD-SD and SD simulations for $T=300$\,K.
Although the difference between the half-widths is negligible for small $q$ values,  
for moderate to large $q$ values, half-widths for the MD-SD results are significantly larger than that for the SD results. 
This indicates significant shortening of the magnon lifetimes due to phonon-magnon scattering.

\subsubsection{Longitudinal magnon modes}

\begin{figure}
 \includegraphics[width=\columnwidth]{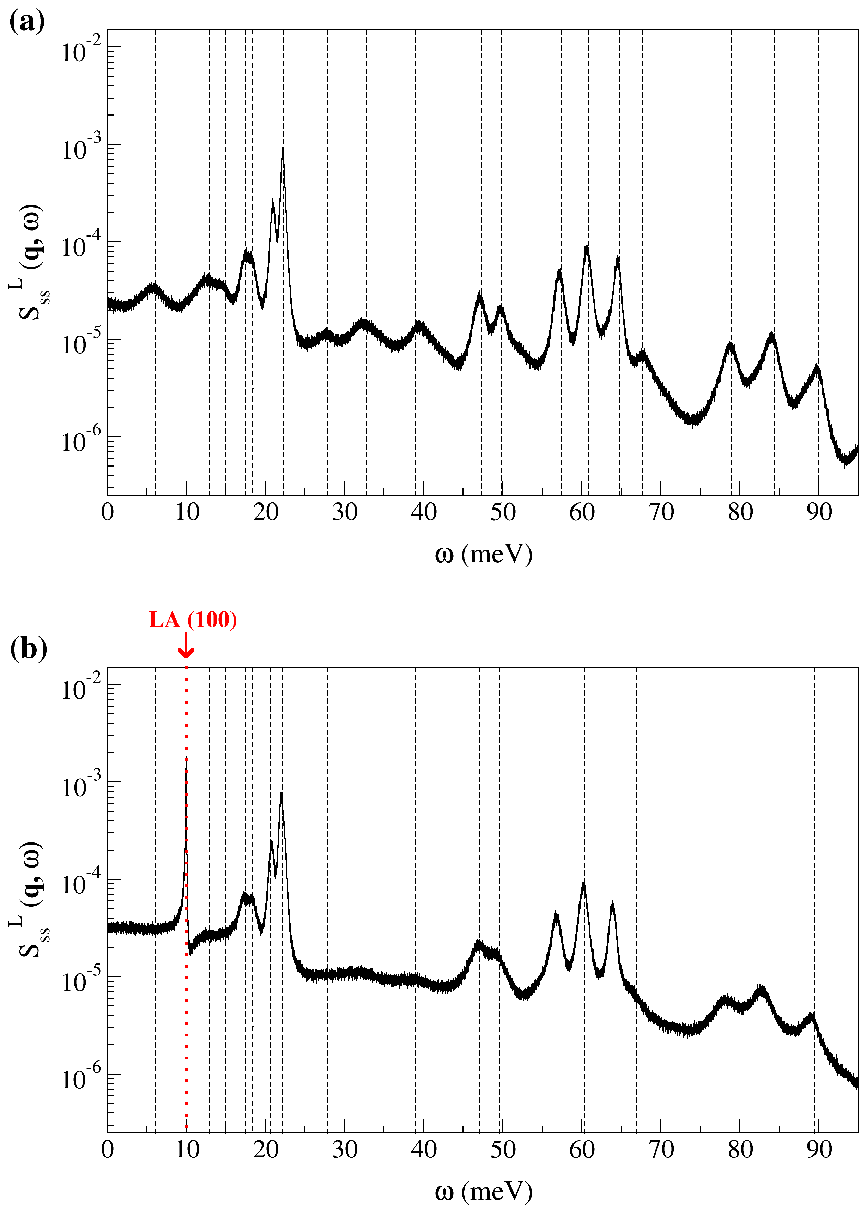}
  \caption{ 
 	The longitudinal component of the spin-spin dynamic structure factor $S_{ss}^L(\mathbf{q}, \omega)$ for 
	$\mathbf{q} = \frac{2 \pi}{La} (1, 0, 0)$ obtained from 
	(a) SD and (b) MD-SD simulations for $L=8$ at $T=300$\,K.
	The predicted positions of the two-spin-wave annihilation peaks are indicated by the vertical dashed lines.
	The dotted line [LA (100)] marks the frequency of the longitudinal phonon mode for the same $\mathbf{q}$.
	  } 
    \label{fig:long_q_1_T_300}
\end{figure}

Our results for the longitudinal spin-spin dynamic structure factor $S_{ss}^L(\mathbf{q}, \omega)$ 
obtained from both MD-SD and SD simulations show many very low-intensity excitations peaks, for all wave vectors considered.
Fig.~\ref{fig:long_q_1_T_300} shows $S_{ss}^L(\mathbf{q}, \omega)$ for a small system size $L=8$ at $T=300$\,K,
where we compare the SD results [panel (a)] with the MD-SD results [panel (b)]
for $\mathbf{q} = \frac{2 \pi}{La}(1, 0, 0)$.

In the context of classical Heisenberg models, Bunker \textit{et al.}~\cite{Bunker2000} showed that 
the excitation peaks observed in $S_{ss}^L(\mathbf{q}, \omega)$ are two-spin-wave creation and/or annihilation peaks
which result from the pairwise interactions between transverse magnon modes.
For ferromagnetic systems, only spin wave annihilation peaks are present, 
and their frequencies are given by
\begin{equation} \label{eq:two_spin_waves}
	\omega_{ij}^- (\mathbf{q}_i \pm \mathbf{q}_j) = \omega(\mathbf{q}_i) - \omega(\mathbf{q}_j),
\end{equation}
where $\mathbf{q}_i$ and $\mathbf{q}_j$ are the wave vectors of the two transverse magnon modes which comprise the two-spin-wave excitation.
Since the set of allowable wave vectors $\{\mathbf{q}_i\}$ depends on the system size $L$,  
the resultant two-spin-wave spectrum also varies with $L$.
For a real magnetic crystal where $L$ is practically infinite, 
the two-spin-wave spectrum would become continuous.

\begin{figure}
 \includegraphics[width=\columnwidth]{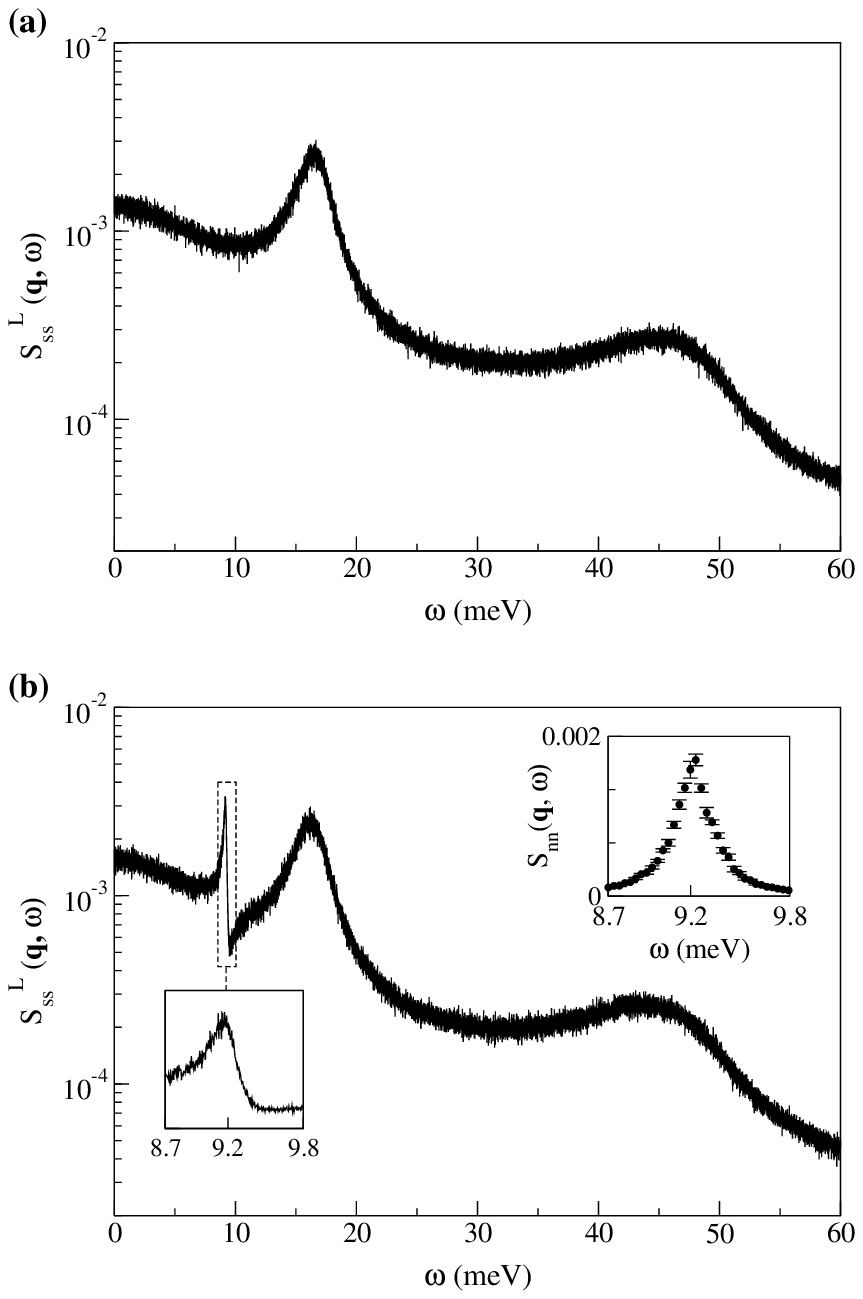}
  \caption{ 
 	The longitudinal component of the spin-spin dynamic structure factor $S_{ss}^L(\mathbf{q}, \omega)$ for 
	$\mathbf{q} = \frac{2 \pi}{La} (1, 0, 0)$ obtained from 
	(a) SD and (b) MD-SD simulations for $L=8$ at $T=800$\,K.
	The inset of (b) shows the density-density dynamic structure factor for the same $\mathbf{q}$.
	  } 
    \label{fig:long_q_1_T_800}
\end{figure}

\begin{figure}
 \includegraphics[width=\columnwidth]{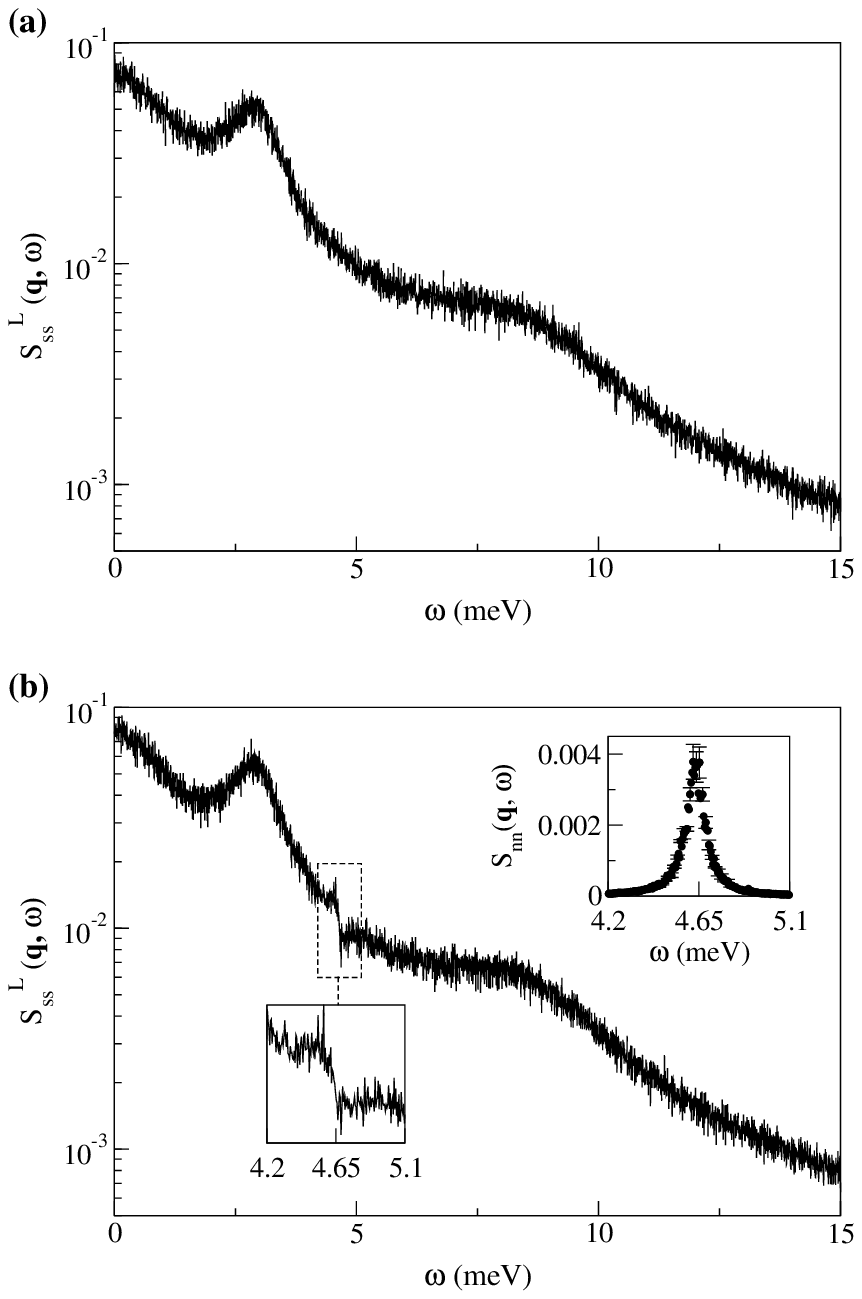}
  \caption{ 
 	The longitudinal component of the spin-spin dynamic structure factor $S_{ss}^L(\mathbf{q}, \omega)$ for 
	$\mathbf{q} = \frac{2 \pi}{La} (1, 0, 0)$ obtained from 
	(a) SD and (b) MD-SD simulations for $L=16$ at $T=1000$\,K.
	The inset of (b) shows the density-density dynamic structure factor for the same $\mathbf{q}$.
	  } 
    \label{fig:long_q_1_T_1000}
\end{figure}

To verify whether the peaks we observe in $S_{ss}^L(\mathbf{q}, \omega)$ are two-spin-wave peaks,
we chose a relatively small system size ($L=8$)
so that the set of allowable wave vectors is reduced to a manageable size.
Then, using MD-SD and SD simulations, we separately determined the transverse magnon frequencies that correspond to the first few $n_q$ values 
along all possible lattice directions.  
With this information at hand, we can predict the expected positions of all the two-spin-wave annihilation peaks using Eq.~\eqref{eq:two_spin_waves}
for both SD and MD-SD case.
As an example, let us consider the wave vector pair $\mathbf{q}_i = (1, 1, 1)$ and $\mathbf{q}_j = (1, 1, 0)$.
Since $\mathbf{q}_i - \mathbf{q}_j = (0, 0, 1)$, they produce a spin wave annihilation peak in $S_{ss}^L(\mathbf{q}, \omega)$ 
for $\mathbf{q} = (0, 0, 1)$ at the frequency  $\omega^- = \omega(\mathbf{q}_i) - \omega(\mathbf{q}_j)$.
(Note that we have ignored the common pre-factor $2 \pi/La$ from the wave vectors.)

In Fig.~\ref{fig:long_q_1_T_300} (a) and (b), we have superimposed the predicted spin wave annihilation peak positions
corresponding to each case.
We see an excellent match between the observed peaks and the predicted two-spin-wave peak positions,
with the exception of the particular sharp peak at $\omega \approx 10$\,meV which only appears in panel (b).
Surprisingly, the position of this peak coincides with the frequency of the longitudinal phonon mode for the same $\mathbf{q}$
as determined from the peak position of $S_{nn}(\mathbf{q}, \omega)$ or $S_{vv}^L(\mathbf{q}, \omega)$.
Similar excitation peaks were observed for all wave vectors, 
for all system sizes and temperatures considered.
The origin of these coupled phonon-magnon modes can be explained as follows.
Unlike transverse phonons, when a longitudinal phonon propagates along a certain lattice direction, it generates fluctuations in the local atom density
along that direction with the corresponding phonon frequency.
This, in turn, leads to fluctuations in the local density of the longitudinal components of the spins (\textit{i.e.} components of the spin vectors
parallel to the net magnetization).
These longitudinal spin fluctuations propagate along with the phonon, yielding a sharp, coupled mode in the the longitudinal magnon spectrum.

Fig.~\ref{fig:long_q_1_T_800} and Fig.~\ref{fig:long_q_1_T_1000} show 
$S_{ss}^L(\mathbf{q}, \omega)$ for $\mathbf{q} = \frac{2 \pi}{La}(1, 0, 0)$ at $T=800$\,K and $T=1000$\,K, respectively,
where we compare the SD results [panel (a)] with the MD-SD results [panel (b)].
In each figure, the inset of panel (b) shows the longitudinal density-density dynamic structure factor for the same wave vector. 
In comparison to the results for $T=300$\,K, we observe that the diffusive central peak becomes more pronounced as the temperature rises,
and many of the low-intensity two-spin-wave peaks broaden and disappear into its tail.
These observations are in qualitative agreement with previous SD studies of the ferromagnetic Heisenberg model~\cite{Bunker2000}.
The coupled phonon-magnon peak also becomes less pronounced with increasing temperature, as the diffusive central peak becomes more pronounced.
At $T=1000$\,K, the intensity of the peak is very low and is barely recognizable.
Above the Curie temperature, spins are randomly oriented and the vector sum of spins per unit volume will be zero on average.  
Hence, the coupled phonon-magnon mode should entirely disappear; 
however, it is already so faint at $T=1000$\,K that we clearly would not have sufficient resolution to test this behavior above the Curie temperature.

We would like to point out that the existence of these coupled phonon-magnon modes is a phenomenon that so far hasn't been discovered experimentally.
In fact, this is not surprising since the experimental detection of these peaks would be extremely challenging
due to their very low intensities.

\section{Conclusions} \label{sec:summary}

To investigate collective phenomena in ferromagnetic bcc iron, we performed combined molecular and spin dynamics (MD-SD) simulations
at temperatures $T=300$\,K, $T=800$\,K, and $T=1000$\,K.
From the trajectories of these simulations, space- and time-displaced correlation functions associated with the atomic and spin variables were calculated.
Fourier transforms of these quantities, namely, dynamic structure factors, directly reveal information regarding the 
frequencies and the lifetimes of the vibrational and magnetic excitation modes.
For small $q$ values, the dispersion relations obtained from our simulations at $T=300$\,K agree well with the experimental results,
but deviations can be observed for large $q$ values, especially for the transverse magnon dispersion curves.
These discrepancies can be attributed to the anharmonic effects not being faithfully captured in the embedded atom potential and the pairwise functional 
representation of the exchange interaction.
In fact, Yin \textit{et al.}~\cite{Yin2012} recently pointed out that the exchange parameters in bcc iron depend on 
the local atomic environment in a complicated manner that may not be properly characterized through a pairwise distance-dependent function.
Thus, a more accurate depiction of magnetic interactions necessitates the development of sophisticated models of exchange interactions 
that effectively capture the contribution of the local environment. 

To understand the mutual influence of the phonons and magnons on each other, 
we compared our results with that of the standalone molecular dynamics and spin dynamics simulations. 
Due to phonon-magnon coupling, we observe a shift in the characteristic frequencies,
as well as a decrease in the lifetimes.
These effects become more pronounced as the temperature is increased.
Moreover, the frequency shifts and the lifetime reductions that occur in magnons due to phonons
are found to be far more pronounced than the corresponding effects experienced by phonons due to magnons. 
This is not surprising considering the fact that the energy scale associated with the spin-spin interactions is about an order of magnitude smaller than 
that of the atomic (non-magnetic) interactions. 

A comparison of our results at different temperatures shows that the effects of spin-lattice coupling becomes more pronounced
as the temperature rises. 
However, due to critical fluctuations, the size of the error bars for magnon properties increases rapidly as the temperature approaches the Curie temperature
(See Fig.~\ref{fig:Sss_T_fitting} (b) and Fig.~\ref{fig:magnon_freq_shift} for example.).
Therefore, obtaining reliable estimates of magnon properties becomes increasingly difficult as the Curie temperature is approached. 

The unprecedented resolution provided by our simulations has allowed us to clearly identify
two-spin-wave peaks in the longitudinal spin-spin dynamic structure factor
with amplitudes down to \textit{six orders of magnitude} smaller than that of the highest single spin wave peak observed.
In addition, in the presence of lattice vibrations, 
we also observe additional longitudinal magnetic excitations with frequencies which coincide with those of the longitudinal phonons.
This is an unexpected form of longitudinal spin wave excitations that so far has not been detected in inelastic neutron scattering experiments,
presumably due to their very low intensities.

\begin{acknowledgments}
This work was sponsored by the ``Center for Defect Physics'', an Energy Frontier Research Center of the Office of Basic Energy Sciences (BES),
U.S. Department of Energy (DOE); the later stages of the work of G.M.S. and M.E. was supported by the Materials Sciences
and Engineering Division of BES, US-DOE. 
We also acknowledge the computational resources provided by the Georgia Advanced Computing Resource Center.
\end{acknowledgments}

\end{document}